\documentclass[aps, pra, reprint, twocolumn, superscriptaddress]{revtex4-2}

\usepackage{graphicx} 
\usepackage{amssymb, amsmath, amsfonts, amsopn, amstext, amsbsy} 
\usepackage{bm} 
\usepackage{color}
\usepackage{float}
\usepackage{setspace}
\usepackage{url}
\usepackage{latexsym}
\usepackage{epstopdf}
\usepackage{threeparttable}
\usepackage{cases}
\usepackage[colorlinks, linkcolor=blue, urlcolor=blue, citecolor=blue, pdfborder={0 0 0}, unicode]{hyperref}

\UseRawInputEncoding

\begin{document}
	
	\title{Higher-order exceptional points and enhanced quantum squeezing \\ in a pseudo-Hermitian semiconductor optomechanical system} 
\author{Zi-Wei Jiang}\thanks{Co-first authors with equal contribution}
\affiliation{Key Laboratory of Low-Dimensional Quantum Structures and Quantum Control of Ministry of Education, Hunan Research Center of the Basic Discipline for Quantum Effects and Quantum Technologies, XJ-Laboratory and Department of Physics, Hunan Normal University, Changsha 410081, China}

\author{Yan-Zi Jing}\thanks{Co-first authors with equal contribution}
\affiliation{Key Laboratory of Low-Dimensional Quantum Structures and Quantum Control of Ministry of Education, Hunan Research Center of the Basic Discipline for Quantum Effects and Quantum Technologies, XJ-Laboratory and Department of Physics, Hunan Normal University, Changsha 410081, China}

\author{Zhen-Sen Lin}
\affiliation{Key Laboratory of Low-Dimensional Quantum Structures and Quantum Control of Ministry of Education, Hunan Research Center of the Basic Discipline for Quantum Effects and Quantum Technologies, XJ-Laboratory and Department of Physics, Hunan Normal University, Changsha 410081, China}

\author{Rui Zhang}
\affiliation{Key Laboratory of Low-Dimensional Quantum Structures and Quantum Control of Ministry of Education, Hunan Research Center of the Basic Discipline for Quantum Effects and Quantum Technologies, XJ-Laboratory and Department of Physics, Hunan Normal University, Changsha 410081, China}

\author{Wen-Quan Yang}
\affiliation{Key Laboratory of Low-Dimensional Quantum Structures and Quantum Control of Ministry of Education, Hunan Research Center of the Basic Discipline for Quantum Effects and Quantum Technologies,  XJ-Laboratory and Department of Physics, Hunan Normal University, Changsha 410081, China}

\author{Ya-Feng Jiao}\email{yfjiao@zzuli.edu.cn}
\affiliation{Academy for Quantum Science and Technology, Zhengzhou University of Light Industry, Zhengzhou 450002, China}

\author{Le-Man Kuang}\email{lmkuang@hunnu.edu.cn}
\affiliation{Key Laboratory of Low-Dimensional Quantum Structures and Quantum Control of Ministry of Education, Hunan Research Center of the Basic Discipline for Quantum Effects and Quantum Technologies,  XJ-Laboratory and Department of Physics, Hunan Normal University, Changsha 410081, China}
	\affiliation{Academy for Quantum Science and Technology, Zhengzhou University of Light Industry, Zhengzhou 450002, China} 
	
\begin{abstract}
We investigate higher-order exceptional points and quantum squeezing of exciton polaritons in a pseudo-Hermitian semiconductor optomechanical system. We show that a third-order exceptional point (EP3) can emerge from the tripartite coupling among photons, excitons, and phonons under pseudo-Hermitian conditions. A pronounced two-mode quantum squeezing of exciton polaritons is revealed, and we demonstrate that this squeezing is significantly enhanced in the vicinity of the EP3. Furthermore, we find that in the PT-symmetric phase, the squeezing dynamics produce a frequency comb of exciton polaritons, whereas the squeezing remains constant over time in the PT-symmetry broken phase and exactly at the EP3. The sudden change in quantum squeezing dynamics can be used to probe the phase transition and the EP3. Our work opens a pathway to manipulate quantum squeezing in semiconductor optomechanical platforms, offering potential advantages for quantum sensing and metrology.
\end{abstract}
	
	\maketitle
	\section{Introduction}
Semiconductor optomechanics \cite{1,2,3,4,5}   explores the coherent interplay among photons, excitons, and mechanical degrees of freedom in engineered semiconductor platforms, establishing a rich intersection of cavity optomechanics, integrated photonics, and solid-state quantum physics. A defining feature of such systems is the strong coupling between confined photons and quantum-well excitons, which yields exciton polaritons—hybrid light–matter quasiparticles with half-light, half-matter character. Owing to their versatile generation, control, and readout capabilities, exciton polaritons have become a powerful testbed for macroscopic quantum phenomena and a promising resource for quantum technologies, including Bose–Einstein condensation \cite{6,7,8,9}, nonclassical light sources \cite{10,11,12,13}, quantum computation \cite{14,15}, quantum simulation \cite{16,17,18,19,20,21}, spin-based switches and memories \cite{22,23}, polaritonic transistors and diodes \cite{24,25}, and nonreciprocal photonic devices \cite{26,27,28}.

Exceptional points (EPs) constitute a fundamental hallmark of non-Hermitian physics. While a generic non-Hermitian Hamiltonian lacks real eigenvalues, operators satisfying the pseudo-Hermiticity condition can exhibit either entirely real spectra or complex-conjugate pairs \cite{29,30,31,32,33,34}.  More intriguingly, non-Hermitian systems can host higher-order exceptional points \cite{35,36,37,38,39,40}, at which three or more eigenmodes coalesce \cite{41,42,43,44,45,46,47,48,49,50}. These higher-order EPs offer fundamentally distinct advantages over EP2s for enhanced sensing \cite{51,52,53,54,55,56}, topological phenomena \cite{57,58}, and entanglement generation \cite{59}. In the context of optomechanics, it has recently been shown that a pseudo-Hermitian cavity optomechanical system comprising three optical and mechanical modes—whether or not it possesses explicit PT symmetry—can support  third-order exceptional points (EP3) \cite{60}, where a quantum phase transition from a PT-symmetric phase (three real eigenenergies) to a PT-broken phase (one real and two complex-conjugate eigenenergies) occurs, accompanied by the simultaneous coalescence of all three eigenmodes.

Despite these advances, higher-order exceptional points remain largely unexplored in pseudo-Hermitian semiconductor optomechanical systems, even though such platforms uniquely combine photonic, excitonic, and phononic degrees of freedom and therefore hold substantial potential for non-Hermitian quantum control and quantum information processing. The investigation of EP3s in this setting is thus highly desirable, as it promises to uncover new quantum phenomena that can be harnessed for continuous-variable quantum technologies.

In this work, we investigate the emergence of a third-order exceptional point and its impact on quantum squeezing in a pseudo-Hermitian semiconductor optomechanical system. We demonstrate that an EP3 arises from the tripartite coherent coupling among photons, excitons, and phonons under pseudo-Hermitian conditions. Notably, we uncover exotic two-mode quantum squeezing of exciton polaritons in the vicinity of the EP3, which exhibits two striking features: (i) a significant enhancement of quantum squeezing, and (ii) the formation of a quantum squeezing frequency comb in the PT-symmetric phase. In contrast, in the PT-broken regime and exactly at the EP3, the squeezing remains constant in time. These results establish a direct connection between higher-order non-Hermitian degeneracies and quantum fluctuations, offering a new route toward EP-enhanced quantum metrology and non-Hermitian continuous-variable quantum information processing.

The paper is organized as follows. In Sec. II, we introduce the physical model of the semiconductor optomechanical system and derive the linearized quantum Langevin equations. Section III is devoted to the analysis of the third-order exceptional point under pseudo-Hermitian conditions. In Sec. IV, we investigate the two-mode squeezing of exciton polaritons and its dependence on the proximity to the EP3. Finally, concluding remarks are given in Sec. V.

	\section{Model AND HAMILTONIAN }

\begin{figure}[htbp]
	\centering
	\includegraphics[width=0.45\textwidth]{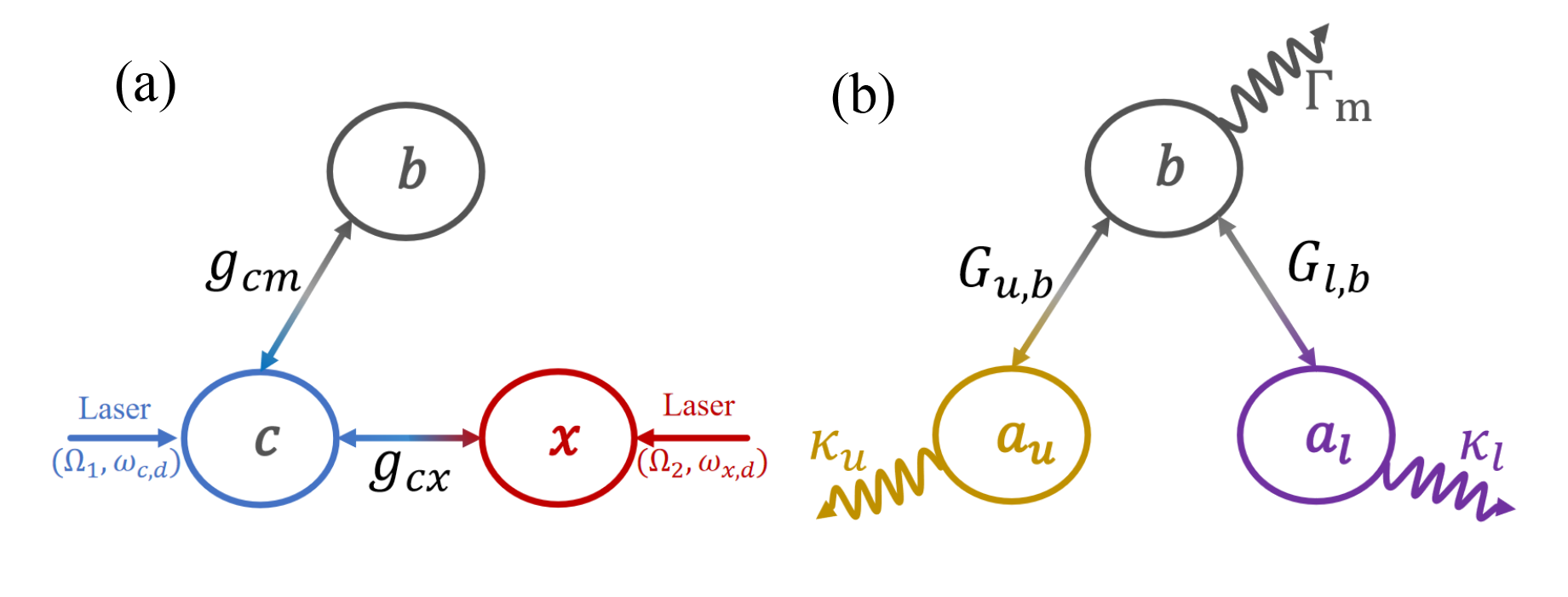}
	\caption{Schematic of the  semicondoctor exciton-optomechanical system. (a)  The tripartite coupling configuration among the cavity mode $c$, the exciton mode $x$, and the mechanical mode $b$. (b) The tripartite coupling configuration among the cavity mode $c$, polariton modes $a_u$ and $a_l$ after linearization in the strong-coupling regime.}
	\label{fig:1}
\end{figure}

We consider a semicondoctor exciton-optomechanical system, which consists of a quantum well (QW) embedded inside a semiconductor microcavity formed by two distributed Bragg reflectors (DBRs). The DBR is a multilayered dielectric structure composed of alternating quarter-wave stacks with distinct refractive indices. By design, the DBR achieves near-perfect reflectivity via constructive interference, thereby forming the cavity mirror. The QW is a thin layer of semiconductor sandwiched between two barrier layers with a much larger band gap. The QW is placed within the microcavity, enabling a linear excitation-exchange (beam-splitter type) interaction between excitons and cavity photons. One of the movable DBRs serves as a mechanical oscillator, which couples to the cavity mode via photon radiation pressure. The exciton-photon-phonon coupled configuration is illustrated in Fig.~\ref{fig:1}(a). The Hamiltonian of the system reads	($ \hbar=1 $)
\begin{equation}
	\label{1}
	\begin{split}
		H = & \omega_{c} c^{\dagger}c + \omega_{x} x^{\dagger}x +\Omega_{m} b^{\dagger}b \\
		& + g_{cm} c^{\dagger}c (b + b^{\dagger}) + g_{cx} (c^{\dagger}x + c x^{\dagger}) \\
		& + \Omega_{1} \left( c e^{i\omega_{c,d}t} + \text{H.c.} \right) + \Omega_{2} \left( x e^{i\omega_{x,d}t} + \text{H.c.} \right),
	\end{split}
\end{equation}
 where $c$, $x$, and $b$ ($c^{\dagger}$, $x^{\dagger}$, and $b^{\dagger}$) denote the annihilation (creation) operators of cavity photons, excitons, and phonons, respectively. These operators satisfy the canonical commutation relation $[j, j^\dagger]=1$ for $j=c,x,b$. The symbols $\omega_{c}$, $\omega_{x}$, and $\Omega_{m}$ represent their respective resonant frequencies. Here, $g_{cm}$ characterizes the optomechanical coupling strength between cavity photons and phonons, while $g_{cx}$ denotes the exciton--photon coupling strength. $\Omega_{1}$ and $\Omega_{2}$ are the driving amplitudes for the photonic and exciton modes, with $\omega_{c,d}$ and $\omega_{x,d}$ as their corresponding driving frequencies.

In the strong-coupling regime, excitons and photons hybridize to form exciton-polaritons, yielding two energy branches: the higher-energy branch corresponds to the upper polariton (UP), and the lower-energy branch corresponds to the lower polariton (LP). Accordingly, the Hamiltonian can be reformulated in the rotating-wave frame at frequency $\omega_d$ by adopting these two polariton modes as follows
\begin{equation}
	\label{2}
	\begin{aligned}
		H_{I} = &\; \Delta_{u} a_{u}^{\dagger} a_{u} + \Delta_{l} a_{l}^{\dagger} a_{l} + \Omega_{m} b^{\dagger} b \\
		& + g_1 a_{u}^{\dagger} a_{u} (b + b^{\dagger}) + g_2 a_{l}^{\dagger} a_{l} (b + b^{\dagger}) \\
		& + g_{h} (a_{u}^{\dagger} a_{l} + a_{l}^{\dagger} a_{u}) (b + b^{\dagger}) \\
		& + (\beta_{u} a_{u} + \beta_{l} a_{l} + \text{H.c.}),
	\end{aligned}
\end{equation}
where $a_{u}$ and $a_{l}$ ($a_{u}^{\dagger}$, $a_{l}^{\dagger}$) denote the annihilation (creation) operators of the upper polariton (UP) and lower polariton (LP), respectively. The exciton ($x$) and cavity photon ($c$) operators are related to the polariton operators ($a_u, a_l$) via a unitary Hopfield transformation:
\begin{equation}
	\label{3}
	a_{u} = c \cos\theta + x \sin\theta ,\quad a_{l} = -c \sin\theta + x \cos\theta,
\end{equation}
where $\theta$ is the mixing angle, defined as $\tan (2\theta) =2g_{cx}/(\omega_c - \omega_x)$. Corresponding to the UP and LP, their resonant frequencies are given by
\begin{equation}
	\label{4}
	\begin{aligned}
		\omega_{u} &= \frac{1}{2}\left[(\omega_{c}+\omega_{x}) + \sqrt{4g_{cx}^{2}+(\omega_{c}-\omega_{x})^{2}}\right],\\
		\omega_{l} &= \frac{1}{2}\left[(\omega_{c}+\omega_{x}) - \sqrt{4g_{cx}^{2}+(\omega_{c}-\omega_{x})^{2}}\right].
	\end{aligned}
\end{equation}

The effective detunings for the UP and LP are defined as $\Delta_{u} = \omega_{u} - \omega_{d}$ and $\Delta_{l} = \omega_{l} - \omega_{d}$, respectively. In the polariton basis, both the polariton-phonon coupling strengths and the driving amplitudes are functions of the mixing angle $\theta$. Specifically, the polariton-phonon coupling strengths are given by  $g_{1} = g_{\text{cm}} \cos^{2}\theta$, $g_{2} = g_{\text{cm}} \sin^{2}\theta$, and $g_{h} = -g_{\text{cm}} \cos\theta \sin\theta$. Meanwhile, the effective driving amplitudes for the UP and LP are $\beta_u = \Omega_1 \cos\theta + \Omega_2 \sin\theta$ and $\beta_l = -\Omega_1 \sin\theta + \Omega_2 \cos\theta$, respectively.
	
Under strong driving conditions, we adopt a linearization approach to handle the Hamiltonian. Each polariton operator is decomposed into the sum of a classical mean value and a quantum fluctuation term, i.e., $a_{u} \rightarrow \alpha_{u} + \delta a_{u}$ and $a_{l} \rightarrow \alpha_{l} + \delta a_{l}$, where $\alpha_{u}$ and $\alpha_{l}$ denote the classical mean values, and $\delta a_{u}$ and $\delta a_{l}$ represent the quantum fluctuation operators. Substituting these decompositions into Eq.~(\ref{2}), neglecting higher-order fluctuation terms, and incorporating the dissipation effect, the linearized quantum equations for the fluctuation operators can be derived as follows:
\begin{equation}
	\label{5}
	\begin{aligned}
		\delta\dot{a}_{u} &= -(i\Delta_{u}+\kappa_{u})\delta a_{u} - i(g_{1}\alpha_{u}+g_{h}\alpha_{l})(b^\dagger + b),\\[6pt]
		\delta\dot{a}_{l} &= -(i\Delta_{l}+\kappa_{l})\delta a_{l} - i(g_{2}\alpha_{l}+g_{h}\alpha_{u})(b^\dagger + b),\\[6pt]
		\dot{b} &= -(i\Omega_{m}+\Gamma_{m})b \\
		&\quad - i\bigl(g_{1}\alpha_{u}^{*}+g_{h}\alpha_{l}^{*}\bigr)\delta a_{u} 
		- i\bigl(g_{1}\alpha_{u}+g_{h}\alpha_{l}\bigr)\delta a_{u}^{\dagger} \\
		&\quad - i\bigl(g_{2}\alpha_{l}^{*}+g_{h}\alpha_{u}^{*}\bigr)\delta a_{l} 
		- i\bigl(g_{2}\alpha_{l}+g_{h}\alpha_{u}\bigr)\delta a_{l}^{\dagger},
	\end{aligned}
\end{equation}
where $\kappa_{u}$ and $\kappa_{l}$ are the decay rates of the upper and lower polariton modes, and $\Gamma_{m}$ is the decay rate of the mechanical mode. The steady-state average values are given by
\begin{equation}
	\label{6}
		\alpha_{u} = \frac{-i\beta_u^*}{i\Delta_u + \kappa_u}, \hspace{0.5cm}
		\alpha_{l} = \frac{-i\beta_l^*}{i\Delta_l + \kappa_l}.
\end{equation}

The linearized Hamiltonian can then be reconstructed from Eq.~(\ref{5}) as
\begin{equation}
	\label{7}
\begin{aligned}
		H_{\text{lin}}  &= (\Delta_{u}-i\kappa_{u})a_{u}^{\dagger}a_{u}+(\Delta_{l}-i\kappa_{l})a_{l}^{\dagger}a_{l}+(\Omega_{m}-i\Gamma_{m})b^{\dagger}b \\ &\quad+G_{u,b}(a_{u}^{\dagger}+a_{u})(b^{\dagger}+b)+G_{l,b}(a_{l}^{\dagger}+a_{l})(b^{\dagger}+b),
	\end{aligned}
\end{equation}
where we have defined the effective coupling strengths as $G_{u,b} = g_1 \alpha_{u} + g_{h} \alpha_{l}$ and $G_{l,b} = g_2 \alpha_{l} + g_{h} \alpha_{u}$. Based on the steady-state solutions, the classical mean values can be made real by appropriately tuning the driving fields.

For simplicity, we have dropped  the $\delta$ symbol for quantum fluctuation operators. In the strong-coupling regime, the linearized model is consistent with Fig.~\ref{fig:1}(b), in comparison with Eq.~(\ref{2}), we can see that the two exciton-polariton modes no longer exhibit direct coupling. Instead, each independently interacts with the mechanical mode.

We consider the case of the opposite-direction detunings with $ \Delta_u> 0 $, $ \Delta_l< 0 $. In this situation,  the rotating-wave approximation is allowed,  fast-oscillating terms in Eq.~(\ref{7}) can be neglected. Then Eq.~(\ref{7}) reduces to
\begin{equation}
	\label{8}
	\begin{aligned}
		H_{\text{eff}} &= (\Delta_{u}-i\kappa_{u})a_{u}^{\dagger}a_{u}+(\Delta_{l}-i\kappa_{l})a_{l}^{\dagger}a_{l}+(\Omega_{m}-i\Gamma_{m})b^{\dagger}b \\
		&\quad + G_{u,b}(a_{u}^{\dagger}b+a_{u}b^{\dagger})+G_{l,b}(a_{l}b+a_{l}^{\dagger}b^{\dagger}).
	\end{aligned}
\end{equation}
In the following, we shall higher-order exceptional points and their influence on the quantum squeezing of the exciton polaritons.

\section{Analyses of higher-order exceptional points}

In this section, we investigate higher-order exceptional points of the pseudo-Hermitian semiconductor optomechanical system under our consideration. In order to do this, we express 
the effective Hamiltonian given in Eq.~(\ref{8}) as the matrix form   $H_{\text{eff}} = \mathbf{X}^\dagger M \mathbf{X}$, where $\mathbf{X} = (a_u, a_{l}^{\dagger}, b)^T$, and $M$ denotes the corresponding coefficient matrix
\begin{equation}
	\label{9}
	M=\begin{pmatrix}\Delta_{u}-i\kappa_{u} & 0 & G_{u,b}\\
		0 & -(\Delta_{l}+i\kappa_{l}) & -G_{l,b}\\
		G_{u,b} & G_{l,b} & \Omega_{m}-i\Gamma_{m}
	\end{pmatrix},
\end{equation}
which is evidently non-Hermitian. 

In generic non-Hermitian systems, searching for exceptional points (EPs) is challenging owing to the high-dimensional parameter space. However, imposing pseudo-Hermitian constraints can greatly reduce the number of independent parameters. In particular, the system described by the effective Hamiltonian in Eq.~(\ref{8}) satisfies pseudo-Hermiticity when its three eigenvalues are either entirely real or composed of one real eigenvalue plus a complex-conjugate pair. Accordingly, we impose the pseudo-Hermitian condition, which requires that
\begin{equation}
	\label{10}
	\det(M - \Omega I) = 0 \quad \text{and} \quad \det(M^{*} - \Omega I) = 0,
\end{equation}
which share the same set of eigenvalues with $\Omega$ denoting an eigenvalue of the effective Hamiltonian $H_{\mathrm{eff}}$ and $I$   the identity matrix. 

By comparing the two characteristic equations, we derive the parameter condition for pseudo-Hermiticity as
	\begin{equation}
			\label{11}
		\begin{aligned}
		\Gamma_{m} &= -(1+\lambda)\kappa_{l}, \\
		(\Delta_{u}- \Omega_{m})\lambda &= (\Delta_{l} + \Omega_{m}), \\
		(G_{u,b}^{2} - \lambda G_{l,b}^{2}) &= \lambda(1+\lambda)\left[(\Delta_{u} - \Omega_{m})^{2} + \kappa_{l}^{2}\right],
	\end{aligned}
	\end{equation}
where the parameter $\lambda = \kappa_{u}/\kappa_{l}$ is introduced to simplify the parametric equations. The system described by Eq. (\ref{8})is pseudo-Hermitian only when the three equations in Eq.~(\ref{11}) are simultaneously satisfied. The first equation reveals the relationship among dissipations of the three modes, indicating that gain and loss must be balanced. The second equation discloses the relation between detunings and dissipation. The third equation shows the interconnection among coupling strengths, dissipation, and detuning.
	
Analyzing the third equation in Eq.~(\ref{11}), it is straightforward to obtain $\lambda(1+\lambda) > 0$,  which implies that $\lambda$ lies in the range
\begin{equation}
	\label{12}
	\lambda > 0 \quad \text{or} \quad \lambda < -1.
\end{equation}

Substituting the pseudo-Hermitian condition Eq.~(\ref{11}) into the characteristic equation $\det(M - \Omega I) = 0$ yields a simplified form, which can be expressed as
	\begin{equation}
			\label{13}
		\Omega^3 + c_2\Omega^2 + c_1\Omega + c_0 = 0,
	\end{equation}
where the three coefficients are given by
	\begin{equation}
			\label{13+}	
		\begin{aligned}
	c_{0} &= (\lambda+1)\Omega_{m} G_{u,b}^{2} - \left(\lambda+\frac{1}{\lambda}\right)\Delta_{u} G_{u,b}^{2}   \\
	&\quad + (\lambda+1)\Delta_{u}(\lambda^{2}\kappa_{l}^{2}+\Delta_{u}^{2}) - (\lambda+2)\Omega_{m}(\lambda^{2}\kappa_{l}^{2}+\Delta_{u}^{2}), \\[6pt]
	c_{1} &= \frac{1-\lambda}{\lambda} G_{u,b}^{2} - (2\lambda+1)\Delta_{u}^{2} + 2(\lambda+2)\Delta_{u}\Omega_{m} + \lambda^{2}\kappa_{l}^{2}, \\[6pt]
	c_{2} &= (\lambda-1)\Delta_{u} - (\lambda+2)\Omega_{m}.
	\end{aligned}
	\end{equation}

According to Cardano's formula, the cubic equation's roots depend on the discriminant
	\begin{equation}
			\label{14}
		\begin{aligned}
		\Delta &= B^2 - 4AC, \\
		A &= c_2^2 - 3c_1, \quad B = c_1c_2 - 9c_0, \quad C = c_1^2 - 3c_0c_2.
		\end{aligned}
    \end{equation}

Whether the roots of the equation (\ref{13}) exhibit degeneracy  is determined by the sign of the discriminant $\Delta$. Depending on the value of $\Delta$, we have the following four cases: (i) if $\Delta > 0$, the equation has one real root and a pair of complex conjugate roots; (ii) if $\Delta < 0$, the equation has three distinct real roots; (iii) if $\Delta = 0$ and $A = B = 0$, the equation has a triple degenerate real root, corresponding to a third-order exceptional point (EP3); (iv) if $\Delta = 0$ and $A \neq 0$, $B \neq 0$, the equation has a double degenerate real root, corresponding to a second-order exceptional point (EP2).
Clearly, the situation we are interested in is the case of $\Delta = 0$, which indicates the emergence of EP3 (triple degeneracy) or EP2 (double degeneracy).
	
We consider the resonant condition $ \Delta_{u}=-\Delta_{l}=\Omega_{m} $ that simplifies  dynamics of the system under our consideration, the third equality of pseudo-Hermitian condition in Eq.~(\ref{11}) simplifies to
	\begin{equation}
		\label{15}
		G_{u,b}^{2}= \lambda G_{l,b}^{2}+\lambda(1+\lambda) \kappa_{l}^{2}.
	\end{equation}
Meanwhile, the second equality in Eq.~(\ref{11}) holds trivially under the above resonant condition.

Substituting the simplified relation in Eq.~(\ref{15}) into the cubic characteristic equation, the coefficients simplify to
\begin{equation}
	\label{16}
	\begin{aligned}
		c_0 &= -\Omega_m\bigl(Q + \Omega_m^2\bigr), \\
		c_1 &= Q + 3\Omega_m^2, \\
		c_2 &= -3\Omega_m,
	\end{aligned}
\end{equation}
where we have introduced the parameter $ Q = (1-\lambda)G_{l,b}^2 + \kappa_l^2 $.

Correspondingly, the discriminant $\Delta$, the intermediate quantities $A$ and $B$  are given by
\begin{equation}
	\label{17}
	\begin{aligned}
	\Delta= 12Q^3, \quad	A = -3Q, \quad B= 6\Omega_m Q,
	\end{aligned}
\end{equation}
which indicates that the condition $\Delta = 0$ yields $Q = 0$, and consequently $A = B = 0$, indicating a triple degeneracy. Thus, only one EP3 emerges under the resonant condition.
	
We now turn to find the EP3 by introducing $ z=\Omega-\Omega_{m} $. In this case, the characteristic equation  reduces to
	\begin{equation}
		\label{18}
	z\left( z^{2}+Q \right)= 0,
	\end{equation}
which admits three eigenvalues for z.After substituting  $ z=\Omega-\Omega_{m} $, the three characteristic roots of the original system read
\begin{equation}
	\label{19}
	\Omega_{0}=\Omega_{m},\quad\Omega_{\pm}=\Omega_{m}\pm\sqrt{-Q}.
\end{equation}

\begin{figure}[htbp]
	\centering
	\includegraphics[width=0.3\textwidth]{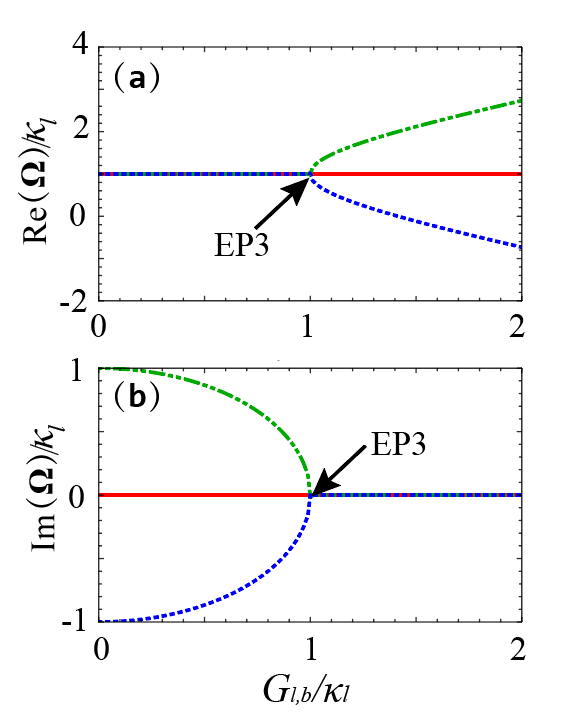}
	\caption{Real and imaginary parts of the eigenvalues as functions of the dimensionless coupling strength $G_{l,b}/\kappa_{l}$.  (a) and (b) show the real and imaginary parts of the eigenvalues, respectively. The solid (red) curve represents $\Omega_{0}$, the dash-dot (green) curve corresponds to $\Omega_{+}$, and the dot (blue) curve corresponds to $\Omega_{-}$. Here, $\lambda = 2$ and $\Omega_{m}/\kappa_{l}=1$.}
	\label{fig:2}
\end{figure}

When $Q < 0$, the system under our consideration possesses three distinct real roots with $\Delta < 0$, which implies that the three eigenmodes of the polariton-optomechanical system are well-separated and exhibit non-degenerate energy levels. When $Q > 0$, there exists one real root $\Omega_{0} = \Omega_{m}$ and a pair of complex-conjugate roots $\Omega_{\pm}$, corresponding to $\Delta > 0$. This indicates that one eigenmode is in resonance with the mechanical mode, while the other two eigenmodes form a paired state with symmetric energy distributions relative to the mechanical resonance frequency. When $Q = 0$, the three roots coalesce completely as $\Omega_{0} = \Omega_{\pm} = \Omega_{m}$, marking the emergence of an EP3 where the three eigenmodes merge into a single degenerate mode.

To clearly illustrate this result, we plot in Fig.~\ref{fig:2} the real and imaginary parts of the eigenvalues $\Omega$ as functions of the dimensionless coupling parameter $G_{l,b}/\kappa_l$ for $\lambda = 2$. When $G_{l,b}/\kappa_l < 1$, $\Omega_0$ remains real while $\Omega_{\pm}$ form a complex-conjugate pair, indicating the PT-broken phase. At $G_{l,b}/\kappa_l = 1$, all three eigenvalues coalesce to $\Omega_{\mathrm{EP3}} = \Omega_m$, marking the emergence of the EP3. For $G_{l,b}/\kappa_l > 1$, the triple degeneracy splits into three distinct real eigenvalues, corresponding to the PT-symmetric phase. Under this choice of parameters ($\lambda = 2$), the explicit condition for the EP3 is $G_{u,b} = 2\sqrt{2}\,\kappa_l$ and $G_{l,b} = \kappa_l$.

Above analyses indicate that under a resonant condition where the detunings of the two polariton modes are equal in magnitude but opposite in sign and both match the mechanical resonance frequency, the characteristic equation simplifies dramatically. The system’s eigenvalue structure is then controlled by a single composite parameter  $G_{l,b}/\kappa_{l}$. When $G_{l,b}/\kappa_{l}>1$, the system lies in the PT‑symmetric phase, featuring three distinct real eigenvalues. When $G_{l,b}/\kappa_{l}<1$, the system enters the PT‑broken phase, with one real eigenvalue and two complex‑conjugate eigenvalues. At the point of $G_{l,b}/\kappa_{l}=$, all three eigenvalues coalesce to the same real value — this marks the emergence of a EP3. The EP3 therefore acts as a phase boundary between the PT‑symmetric and PT‑broken regimes. We explicitly demonstrate, using a representative parameter set, that the transition occurs at a specific coupling strength. Importantly, the system under consideration hosts only a single EP3, at which all three eigenmodes become degenerate simultaneously. This analysis lays the foundation for understanding how the EP3 influences quantum fluctuations, particularly the two‑mode squeezing of exciton polaritons examined later in the following section.

\section{Two-mode squeezing of exciton polaritons}

In this section, we study two-mode squeezing of exciton polaritons and explore novel phenomenon  in induced by the EP3  under resonant conditions. Since quantum squeezing is characterized by the quantum fluctuations of the system operators, we start from the Heisenberg equations of motion to analyze the dynamic evolution of these fluctuations. The Heisenberg equations derived from Eq.~(\ref{8}) can be written as
	\begin{equation}
		\label{21}
		\begin{aligned}\frac{d}{dt}a_{u} & =-i(\Omega_{m}-i\kappa_{u})a_{u}-iG_{u,b}b,\\
			\frac{d}{dt}a_{l}^{\dagger} & =i(-\Omega_{m}+i\kappa_{l})a_{l}^{\dagger}+iG_{l,b}b,\\
			\frac{d}{dt}b & =-i(\Omega_{m}-i\Gamma_{m})b-iG_{u,b}a_{u}-iG_{l,b}a_{l}^{\dagger}.
		\end{aligned}
	\end{equation}

In order to simplify the solution process, we introduce the slowly varying operators $A_u(t)=e^{i\Omega_m t}a_u(t)$, $A_l^\dagger(t)=e^{i\Omega_m t}a_l^\dagger(t)$ and $B(t)=e^{i\Omega_m t}b(t)$. By performing the Laplace transform, we obtain the analytical solutions for the operators 
	  \begin{equation}
	  	\label{22}
	  	\begin{aligned}
	  		a_u(t) &= e^{-i\Omega_m t}\left[f_1 a_u(0) + f_2 a_l^\dagger(0) + f_3 b(0)\right] ,\\[6pt]
	  		a_l^\dagger(t) &= e^{-i\Omega_m t}\left[-f_2 a_u(0) + f_4 a_l^\dagger(0) + f_5 b(0)\right], \\[6pt]
	  		b(t) &= e^{-i\Omega_m t}\left[f_3 a_u(0) - f_5 a_l^\dagger(0) + f_6 b(0)\right]，
	  	\end{aligned}
	  \end{equation}
where we have used the pseudo-Hermitian constraint conditions, the coefficients are given by
 \begin{equation}
 	\label{23}
 	\begin{aligned}
 		f_{1}(t) & =\cosh\omega t-\frac{\lambda\kappa_{l}}{\omega}\sinh\omega t-\frac{G_{u,b}^{2}(\cosh\omega t-1)}{\lambda\omega^{2}},\\
 		f_{2}(t) & =-\frac{G_{u,b}G_{l,b}(\cosh\omega t-1)}{\omega^{2}},\\
 		f_{3}(t) & =-iG_{u,b}\left[\frac{\sinh\omega t}{\omega}+\frac{\kappa_{l}(\cosh\omega t-1)}{\omega^{2}}\right],\\
 		f_{4}(t) & =\cosh\omega t-\frac{\kappa_{l}}{\omega}\sinh\omega t+\frac{\lambda G_{l,b}^{2}(\cosh\omega t-1)}{\omega^{2}},\\
 		f_{5}(t) &=iG_{l,b}\left[\frac{\sinh\omega t}{\omega}+\frac{\lambda\kappa_{l}(\cosh\omega t-1)}{\omega^{2}}\right],\\
 		f_{6}(t) & =\cosh\omega t+\frac{(1+\lambda)\kappa_{l}}{\omega}\sinh\omega t+\frac{\lambda\kappa_{l}^{2}(\cosh\omega t-1)}{\omega^{2}}，
 	\end{aligned}	
 \end{equation}
where  we have introduced the frequency $\omega = \sqrt{\kappa_l^2 - (\lambda-1)G_{l,b}^2}$.
 
 Eq.~(\ref{23}) shows that for real $\omega$, the operator solutions are expressed in terms of hyperbolic functions; for purely imaginary $\omega$, they reduce to trigonometric functions. At the exceptional point $\omega=0$, the coefficients $f_i(t)$ take the degenerate limiting form
 \begin{equation}
 	\label{24}
 	\begin{aligned}
 		f_{1,\text{ep}}(t) &= 1 - \lambda\kappa_l t - \frac{1}{2}\lambda^2 G_{l,b}^2 t^2, \\
 		f_{2,\text{ep}}(t) &= -\frac{1}{2}G_{u,b}G_{l,b}t^2, \\
 		f_{3,\text{ep}}(t) &= -iG_{u,b}\left[t + \frac{\kappa_l}{2} t^2\right], \\
 		f_{4,\text{ep}}(t) &= 1 - \kappa_l t + \frac{1}{2}\lambda G_{l,b}^2 t^2, \\
 		f_{5,\text{ep}}(t) &= iG_{l,b}\left[t + \frac{\lambda\kappa_l}{2} t^2\right], \\
 		f_{6,\text{ep}}(t) &= 1 + (1+\lambda)\kappa_l t + \frac{1}{2}\lambda\kappa_l^2 t^2.
 	\end{aligned}
 \end{equation}
 
 The time-dependent coefficients $f_i(t)$ adopt different forms determined by the system parameters. Away from the EP3 with $\omega \neq 0$, the solutions are described by Eq.~(\ref{23}); exactly at the EP3 with $\omega = 0$, they reduce to the degenerate forms $f_{i,\text{ep}}(t)$ in Eq.~(\ref{24}). These coefficients provide a fundamental basis for investigating the quantum squeezing behavior in the vicinity of the EP3.
	
To investigate the two-mode squeezing effect of exciton polaritons, we introduce two pairs of quadrature operators for the polariton modes $a_u$ and $a_l$ defined by
\begin{equation}
	\label{30}
	\begin{aligned}
		X_{a_u} &= \frac{1}{\sqrt{2}}(a_u + a_u^\dagger), \quad X_{a_l} = \frac{1}{\sqrt{2}}(a_l + a_l^\dagger), \\
		Y_{a_u} &= \frac{-i}{\sqrt{2}}(a_u - a_u^\dagger), \quad Y_{a_l} = \frac{-i}{\sqrt{2}}(a_l - a_l^\dagger).
	\end{aligned}
\end{equation}

We then construct the collective mode operator $C$ of two polariton modes as
\begin{equation}
	\label{31}
	C = \frac{1}{\sqrt{2}}(a_u + a_l),
\end{equation}
which leads to the quadrature operators of the collective mode $C$ as 
\begin{equation}
	\label{32}
	X_C = \frac{1}{\sqrt{2}}(X_{a_u} + X_{a_l}), \quad 
	Y_C = \frac{1}{\sqrt{2}}(Y_{a_u} + Y_{a_l}),
\end{equation}
which  yields the Heisenberg uncertainty relation
\begin{equation}
	\label{26}
	\langle(\Delta X_C)^{2}\rangle\langle(\Delta Y_C)^{2}\rangle\geq\frac{1}{4}|\langle[X_C,Y_C]\rangle|^{2},
\end{equation}
where the variances are defined as $\langle(\Delta X_C)^{2}\rangle = \langle X_C^2 \rangle - \langle X_C \rangle^2$ and $\langle(\Delta Y_C)^{2}\rangle = \langle Y_C^2 \rangle - \langle Y_C \rangle^2$.

The quantum squeezing  of one quadrature emerges if one of the following inequalities holds
\begin{equation}
	\label{27}
	\langle(\Delta X_C)^{2}\rangle < \frac{1}{2}|\langle[X_C, Y_C]\rangle|   \text{or}   \langle(\Delta Y_C)^{2}\rangle < \frac{1}{2}|\langle[X_C, Y_C]\rangle|.
\end{equation}

For a quantitative characterization of quantum squeezing, we introduce the time-dependent squeezing parameters 
\begin{equation}
	\label{28}
	\begin{aligned}
		S_{X_C}&=\frac{\langle(\Delta X_C(t))^{2}\rangle-\dfrac{1}{2}|\langle[X_C(t),Y_C(t)]\rangle|}{\dfrac{1}{2}|\langle[X_C(t),Y_C(t)]\rangle|},\\
	S_{Y_C}&=\frac{\langle(\Delta Y_C(t))^{2}\rangle-\dfrac{1}{2}|\langle[X_C(t),Y_C(t)]\rangle|}{\dfrac{1}{2}|\langle[X_C(t),Y_C(t)]\rangle|}.
	\end{aligned}
\end{equation}
Accordingly, the squeezing condition takes the simple form
\begin{equation}
	\label{29}
	S_{X_C} < 0 \quad \text{or} \quad S_{Y_C} < 0.
\end{equation}
Since the quantum variance is non-negative, the squeezing parameter attains its theoretical lower bound $-1$ as the variance approaches zero. Hence, $S_{X_C}=-1$ ($S_{Y_C}=-1$) means that the $X_C$  ($Y_C$) quadrature reaches the maximum squeezing.

Taking the initial state as the vacuum state $|0\rangle$ and employing the operator solutions in Eq.~(\ref{22}), we derive the variances and the mean value of the commutator  for the collective mode $C$ as
\begin{equation}
	\label{33}
	\begin{aligned}
		\langle(\Delta X_{C})^{2}\rangle 
		&= \frac{1}{4}\big(|f_{1}-f_{2}|^{2} + |f_{2}+f_{4}|^{2} + |f_{3}+f_{5}|^{2}\big),\\
		\langle(\Delta Y_{C})^{2}\rangle &=\frac{1}{4}\big(|f_{1}+f_{2}|^{2} + |f_{2}-f_{4}|^{2} + |f_{3}-f_{5}|^{2}\big),\\
		\langle[X_{C},Y_{C}]\rangle &=\frac{1}{2}\left(|2|f_{2}|^{2}+|f_{5}|^{2}-|f_{1}|^{2}-|f_{3}|^{2}-|f_{4}|^{2}\right).
	\end{aligned}
\end{equation}

\begin{figure}[htbp]
	\centering
	\includegraphics[width=0.50\textwidth]{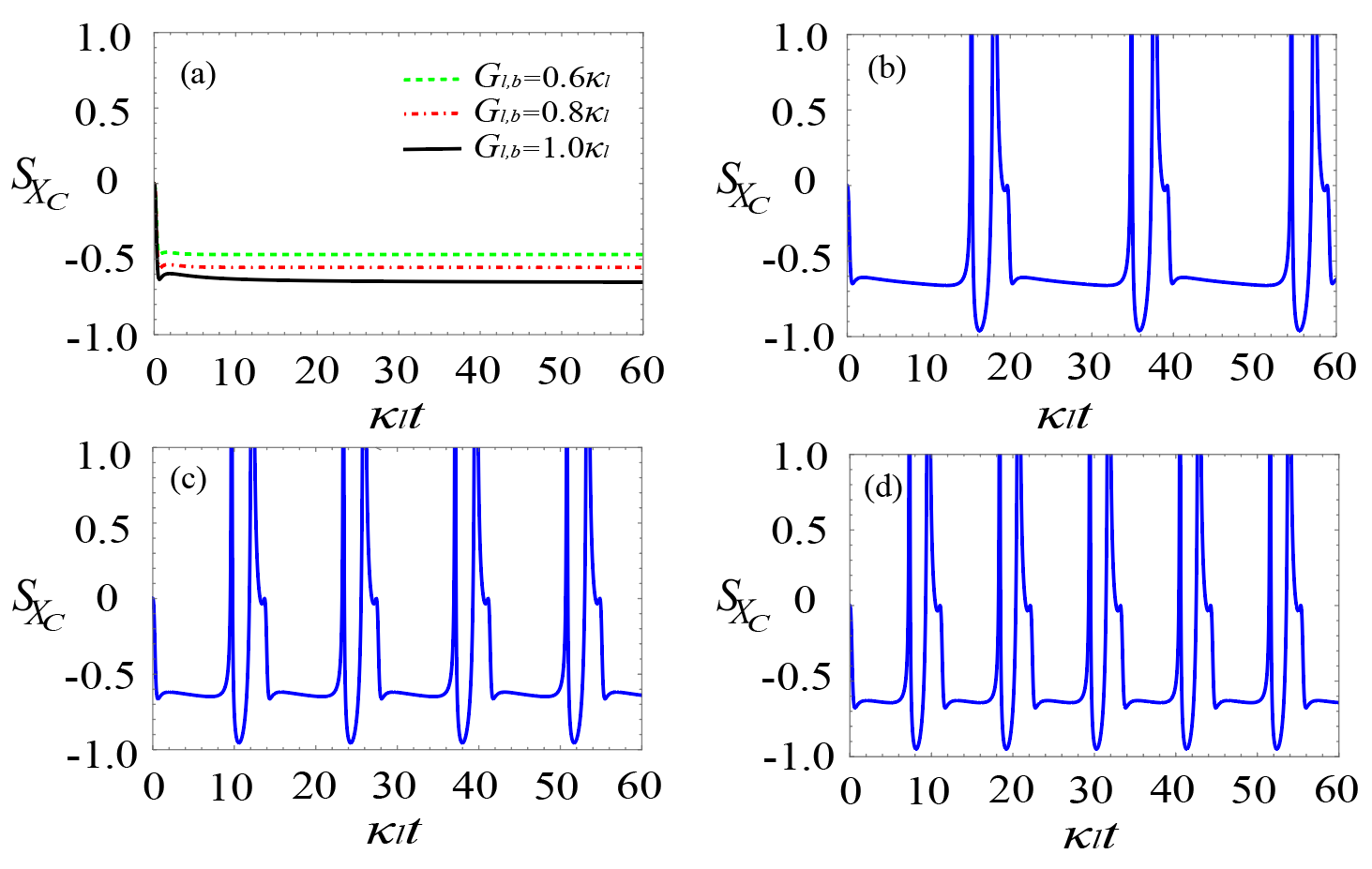}
	\caption{ Dynamic evolution of the quantum squeezing described by the squeezing parameter $S_{X_C}$ versus the dimensionless time $\kappa_l t$ for typical coupling strengths. 
		(a) Cases of PT-symmetry broken phase ($G_{l,b}/\kappa_l = 0.6$, $0.8$) and EP3 ($G_{l,b}/\kappa_l = 1.0$);
		(b)-(d) Cases of PT-symmetric phase for $G_{l,b}/\kappa_l = 1.05$, $1.10$, and $1.15$, respectively. Here we take $\lambda = 2$.}
	\label{fig:3}
\end{figure}

The squeezing parameters $S_{X_C}$ and $S_{Y_C}$ for the two quadrature components of the collective mode $C$  then can be calculated with the following expressions
\begin{equation}
	\label{34}
	\begin{aligned}
		S_{X_{C}} & = \frac{|f_{1}-f_{2}|^{2}+|f_{2}+f_{4}|^{2}+|f_{3}+f_{5}|^{2}}{\left|2|f_{2}|^{2}+|f_{5}|^{2}-|f_{1}|^{2}-|f_{3}|^{2}-|f_{4}|^{2}\right|} - 1, \\[10pt]
		S_{Y_{C}} & =  \frac{|f_{1}+f_{2}|^{2}+|f_{2}-f_{4}|^{2}+|f_{3}-f_{5}|^{2}}{\left|2|f_{2}|^{2}+|f_{5}|^{2}-|f_{1}|^{2}-|f_{4}|^{2}-|f_{3}|^{2}\right|} - 1.
	\end{aligned}
\end{equation}

To explicitly illustrate how the third-order exceptional point (EP3) modulates two-mode squeezing, we numerically investigate the evolution of the squeezing parameter $S_{X_C}$ as a function of the dimensionless time $\kappa_l t$ for several typical coupling strengths. Fig.~\ref{fig:3}(a) presents the dynamical evolution of $S_{X_C}$ in the PT-symmetry-broken phase and at the EP3. The squeezing parameter drops rapidly within an extremely short time and then remains nearly constant; a larger coupling strength yields a stronger squeezing effect. Figures~\ref{fig:3}(b)-(d) correspond to the PT-symmetric phase, where the squeezing parameter displays pronounced oscillatory behavior and forms a squeezing comb. Near the comb troughs the values approach the theoretical lower bound of $-1$, indicating strong instantaneous squeezing. Moreover, the spacing between the comb teeth can be tuned by adjusting the coupling strength between the lower polariton and the mechanical mode, with a larger coupling strength leading to a smaller tooth spacing.

These results demonstrate that the PT-symmetric and PT-symmetry-broken phases exhibit distinct squeezing characteristics. Their unique dynamical features, which arise from the different analytical solutions, become clearly visible only after long-time evolution. Accordingly, in the subsequent analysis, we fix the evolution time and adopt the coupling strength as a variable to explore the squeezing boundary characteristics of the different phases.

\begin{figure}[htbp]
	\centering
	\includegraphics[width=0.50\textwidth]{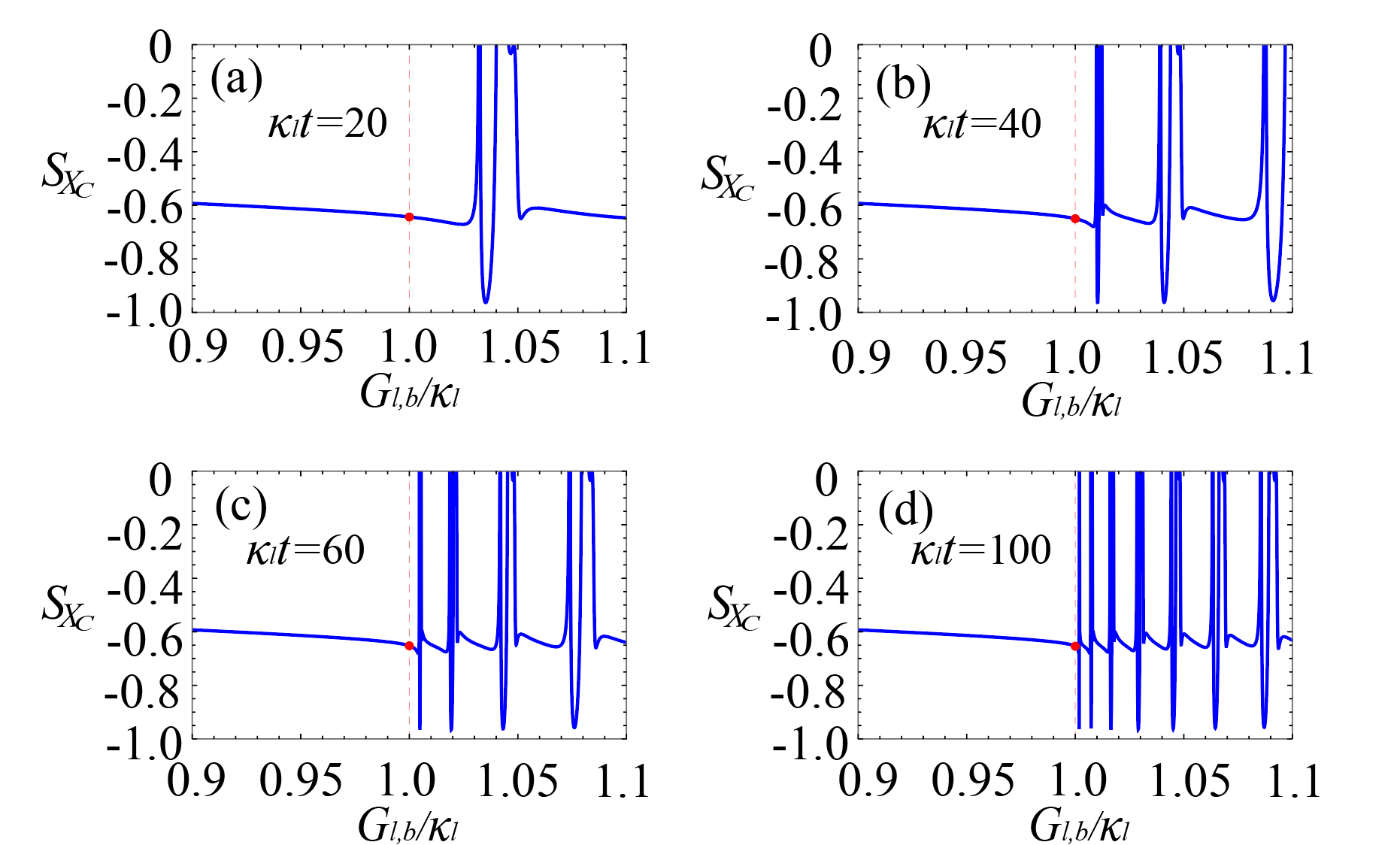}
	\caption{The quantum squeezing described by the squeezing parameter $S_{X_C}$ versus the dimensionless coupling strength $G_{l,b}/\kappa_l$ at different dimensionless evolution times 
		(a) $\kappa_l t=20$; (b) $\kappa_l t=40$; (c) $\kappa_l t=60$; (d) $\kappa_l t=100$. Here we take $\lambda = 2$. The red dot marks the EP3 at $G_{l,b}/\kappa_l = 1$.}
	\label{fig:4}
\end{figure}

We further investigate the dependence of the squeezing parameter $S_{X_C}$ on the dimensionless coupling strength $G_{l,b}/\kappa_l$ at long evolution times, as shown in Fig.~\ref{fig:4}. Several typical instants are selected: $\kappa_l t = 20,\,40,\,60,\,100$. One observes that the squeezing parameter exhibits distinct boundary behaviors across the different PT phases. In the PT-symmetric phase, a clear comb-like structure---referred to as the squeezing comb---gradually emerges. This structure originates from the analytical forms of the time-dependent operator solutions. In the PT-symmetry-broken phase ($G_{l,b}/\kappa_l < 1$), $S_{X_C}$ is governed by hyperbolic functions and gives rise to smooth steady-state squeezing. In the PT-symmetric phase ($G_{l,b}/\kappa_l > 1$), $S_{X_C}$ is dominated by trigonometric functions and displays periodic transient strong squeezing.

At $\kappa_l t = 20$ [Fig.~\ref{fig:4}(a)], the PT-symmetry-broken phase ($G_{l,b}/\kappa_l < 1$) stabilizes near $-0.6$, while the PT-symmetric phase ($G_{l,b}/\kappa_l > 1$) shows a single sharp dip close to $-1$. The first transition appears at $G_{l,b}/\kappa_l \approx 1.03$. At $\kappa_l t = 40$ [Fig.~\ref{fig:4}(b)], the PT-symmetry-broken phase remains nearly unchanged. By contrast, three sharp dips appear in the PT-symmetric phase, and the first transition point shifts leftward to about $1.01$. At $\kappa_l t = 60$ [Fig.~\ref{fig:4}(c)], the squeezing comb becomes more pronounced: the number of comb teeth increases and their spacing gradually decreases. The PT-symmetry-broken phase stays stable at $\sim -0.6$. At $\kappa_l t = 100$ [Fig.~\ref{fig:4}(d)], a dense squeezing comb consisting of numerous sharp dips is formed, and the position of the first comb tooth moves extremely close to $G_{l,b}/\kappa_l = 1.0$. The PT-symmetry-broken phase maintains a steady squeezing near $-0.6$, while all dips in the PT-symmetric phase retain their depths close to $-1$.

It is evident that as the evolution time increases, the first transition point in the PT-symmetric phase gradually approaches the EP3, and the spacing between adjacent comb teeth decreases monotonically. Moreover, at a fixed evolution time, the tooth spacing also becomes smaller as the coupling strength approaches the EP3. Finally, it should be mentioned that the $Y_C$ quadrature of the composite exciton polaritons does not exhibits quantum squeezing in the parameter regimes under our consideration, hence we do not present related analyses on quantum squeezing of the $Y_C$ quadrature.

\section{CONCLUDING REMARKS}

In summary, we have theoretically investigated higher-order exceptional points and quantum squeezing of exciton polaritons in a pseudo-Hermitian semiconductor optomechanical system. By engineering the tripartite coupling among photons, excitons, and phonons, we have demonstrated that a third-order exceptional point  naturally emerges under pseudo-Hermitian conditions, at which a spontaneous PT-symmetry-breaking transition occurs and all three eigenmodes simultaneously coalesce.

Focusing on the system’s quantum fluctuation properties, we have uncovered a pronounced two-mode squeezing of exciton polaritons that is significantly enhanced in the vicinity of the EP3. Notably, the squeezing dynamics exhibit fundamentally distinct behaviors in different phases: in the PT-symmetric phase, the squeezing evolves into a frequency comb structure of exciton polaritons, whereas in the PT-broken phase and exactly at the EP3, the squeezing remains constant over time. The sudden change in quantum squeezing dynamics between the PT-symmetric and PT-symmetry-broken phases can be used to probe the phase transition and the EP3.

These results establish a direct connection between higher-order non-Hermitian degeneracies and quantum correlations, highlighting that EP3s can serve as a powerful resource for manipulating quantum fluctuations in semiconductor optomechanical platforms. Our work thus opens a promising pathway for EP-enhanced quantum sensing and metrology and suggests that pseudo-Hermitian semiconductor optomechanical systems may become a versatile building block for non-Hermitian continuous-variable quantum information processing. Future extensions of this work could explore the experimental realization of higher-order EPs in semiconductor microcavities, and investigate their potential for generating and distributing entangled states in non-Hermitian quantum networks.

	\begin{acknowledgements}
		
		This work is supported by the Natural Science Foundation of China (NSFC) (Grant Nos. 12247105, 12421005, 12405029), the Quantum Science and Technology-National Science and Technology Major Project (Grant No. 2024ZD0301000),  Hunan provincial sci-tech program (Grant No. 2023ZJ1010),  Henan Science and Technology Major Project (No. 241100210400), and  XJ-Lab key project (Grant No. 23XJ02001). 
		
	\end{acknowledgements}

\end{document}